# 4 MV/cm (010) β-$Ga_2O_3$ Heterojunction Diodes Realized by Low-Damage e-Beam $NiO_x$ Interlayers

Carl Peterson,[1,a)] Yizheng Liu,[1] Chinmoy Nath Saha,[1] Rachel Kahler,[1] Akhila Mattapalli,[1] and Sriram Krishnamoorthy[1,a)]

[1]*Materials Department, University of California Santa Barbara, Santa Barbara, California, 93106, USA*

[a)] Author to whom correspondence should be addressed. Electronic mail: carlpeterson@ucsb.edu and sriramkrishnamoorthy@ucsb.edu

We report on the utilization of a low-damage e-beam $NiO_x$ deposition process to realize field-plated heterojunction diodes (FP-HJDs) on (010) $\beta$-$Ga_2O_3$ films with high critical breakdown field strengths beyond 4 MV/cm and power figure of merits (PFOM) of >1 GW/cm$^2$. Diodes were fabricated on a 2.67 × $10^{16}$ cm$^{-3}$ intentionally doped 6.2 μm thick epitaxial layer grown by metalorganic chemical vapor deposition (MOCVD) on a conductive Sn-doped $\beta$-$Ga_2O_3$ (010) substrate using TMGa, $O_2$, Ar carrier gas, and $SiH_4$ as the silicon dopant source. The mesa etched FP-HJD devices utilized a thin 7 nm e-beam $NiO_x$ interlayer before the sputtered $NiO_x$ layers to eliminate the effects of sputter-induced ion damage on the (010) epilayers. Current-Voltage measurements resulted in a forward current density of 700 A/cm$^2$ at 4 V, HJD ideality factor of 1.29, a $V_{bi}$ of 2 V, rectification ratio of $10^{11}$, and a differential specific on resistance ($R_{on,sp}$) value of 2.36 mΩ·cm$^2$. Breakdown of the (010) FP-HJDs was 1.64 kV, leading to a parallel plane electric field at breakdown ($E_{||,max}$) of 4.02 MV/cm and a PFOM of 1.14 GW/cm$^2$, which is a state-of-the-art result for diodes on MOCVD-grown (010) drift layers.

As the emergence of energy-intensive infrastructure such as AI data centers continues to rise, solutions to mitigate the environmental impact of these structures are necessary. An important feature of such infrastructure is power conversion, which comes with inherent resistive losses. Utilizing wide-bandgap (WBG) and ultra-wide band gap (UWBG) semiconductor materials such as SiC, GaN, or beta gallium oxide ($\beta$-$Ga_2O_3$) has shown potential to minimize the size, weight, and power consumption (SWaP) of these high-voltage power conversion applications.[1,2] Recently, extensive research has gone into developing $\beta$-$Ga_2O_3$ power devices due to the combination of desirable properties such as a large bandgap, critical electric field strength of 8MV/cm (or up to 12.9 MV/cm),[3] availability of multiple shallow donors, melt-grown conductive and insulating bulk substrates with dopant impurity control,[4–10] mobility of 200 $cm^2$/Vs,[11] and best-in-class figure of merit among its WBG and UWBG peers[12,13].

To capitalize on $\beta$-$Ga_2O_3$'s unique material properties, the development of high-quality thick epitaxial drift layers and novel high voltage power device architectures must be achieved. The stand-out growth technologies for creating thick low-doped drift layers are halide vapor-phase epitaxy (HVPE),[14–18] metal organic chemical vapor deposition (MOCVD),[11,19–34] and low pressure chemical vapor deposition (LPCVD).[35–38] In particular, MOCVD epilayers are promising as they exhibit the highest electron mobilities among the growth techniques[11] and provide a basis for high-performance power devices.[39–50] For novel power device architecture, sputtered $NiO_x$ heterojunction diodes (HJDs) are of particular interest over conventional Schottky barrier diodes as they have achieved impressive multi-kV-class breakdown voltages and reverse bias leakage currents less than the measurement equipment noise limit.[51–60] However, recent reports show the (010) plane of $\beta$-$Ga_2O_3$ is highly susceptible to ion damage from the conventionally used $NiO_x$ sputtering process,[61] and thus, creation of low-resistance multi-kV HJDs on (010) epilayers has proven difficult.[27,61] In this work, we propose and demonstrate the use of thin, low-damage, e-beam deposited $NiO_x$ interlayers as a successful method for protecting the (010) surface from ion damage during subsequent $NiO_x$ sputtering, enabling low on-resistance and high breakdown field strength $NiO_x$ HJDs on (010) epitaxial layers.

Growth of the 6.2 $\mu$m (010) $\beta$-$Ga_2O_3$ epitaxial drift layer was done using an Agnitron Agilis 100 cold wall MOCVD reactor equipped with a remote injection showerhead. The gallium precursor used for growth was trimethylgallium (TMGa), ultra-high purity $O_2$ gas (5N) was chosen as the oxygen source, dilute silane ($SiH_4$) was used as the silicon source, and high purity Argon was chosen as the carrier gas. The homoepitaxial growth was performed on a 5×5 $mm^2$ Sn-doped conductive (010) $\beta$-$Ga_2O_3$ substrate with no intentional miscut, commercially acquired from NCT. Prior to growth, the substrate was cleaned using acetone, methanol, and deionized (DI) water. Growth was performed in the mass-transport limited regime at a temperature of 1000 ºC with a growth chamber pressure of 15 Torr. Additionally, for all samples the $O_2$, TMGa, $SiH_4$, and Ar flow rates were fixed at 0.089 mol/min, 170.59 µmol/min, 47.83 pmol/min, and 1500 sccm respectively, leading to a VI/III ratio of 523. These reactor conditions led to a growth rate of 3.7 $\mu$m/hr., verified by cross-sectional SEM on a co-loaded c-plane sapphire substrate.[11] The aforementioned growth conditions for thick TMGa drift layers were optimized in a previous work.[33]

Fabrication of field-plated heterojunction diodes (FP-HJDs) on the 6.2 $\mu$m (010) $\beta$-$Ga_2O_3$ epitaxial drift layer began with e-beam deposition of a Ti/Au Ohmic contact on the back side of the Sn-doped substrate, followed by a 470 °C anneal in $N_2$ gas to improve the Ohmic characteristics.[62,63] Next, formation of the anode contacts began with the patterning of a photoresist (PR) lift-off mask defined by optical lithography. After PR development, a thin 7 nm $NiO_x$ layer was deposited via a low-damage e-beam process to act as a protective barrier for the subsequent layers of sputtered $NiO_x$. The sample was then moved quickly from the e-beam to the sputtering chamber to minimize contamination, and a self-aligned 9 nm $p^-$ $NiO_x$/20 nm $p^+$ $NiO_x$ stack was reactively sputtered using an elemental Ni target and a 150 W $O_2$/Ar plasma without breaking vacuum. Changing the resistivity of the $p^-$ ($R_{sheet}$ ~ 215 kΩ/□) and $p^+$ ($R_{sheet}$ ~ 7–9 kΩ/□) $NiO_x$ layers was done by varying the $O_2$/Ar ratio.[58,64] On top of the $NiO_x$, a self-aligned Ohmic Ni/Au/Ni metal stack was deposited via e-beam evaporation to be used as the Ohmic anode metal contact and dry etch mask. After metal deposition and anode contact lift-off, a 2 µm-deep self-aligned heterojunction isolation etch was performed via an inductively coupled plasma (ICP) etch with $BCl_3$ process gas and 30/200 W bias/ICP power. A 1 µm thick $SiO_2$ layer was then reactively sputtered using an elemental Si target at 245 W to act as the field plate dielectric, the $SiO_2$ was patterned via a photoresist lift-off technique. Finally, a lithographically defined Ni/Au field plate (FP) metal was deposited via e-beam with a FP overlap of 8 µm from the edge of the anode. The finalized device structure is shown in FIG. 1.

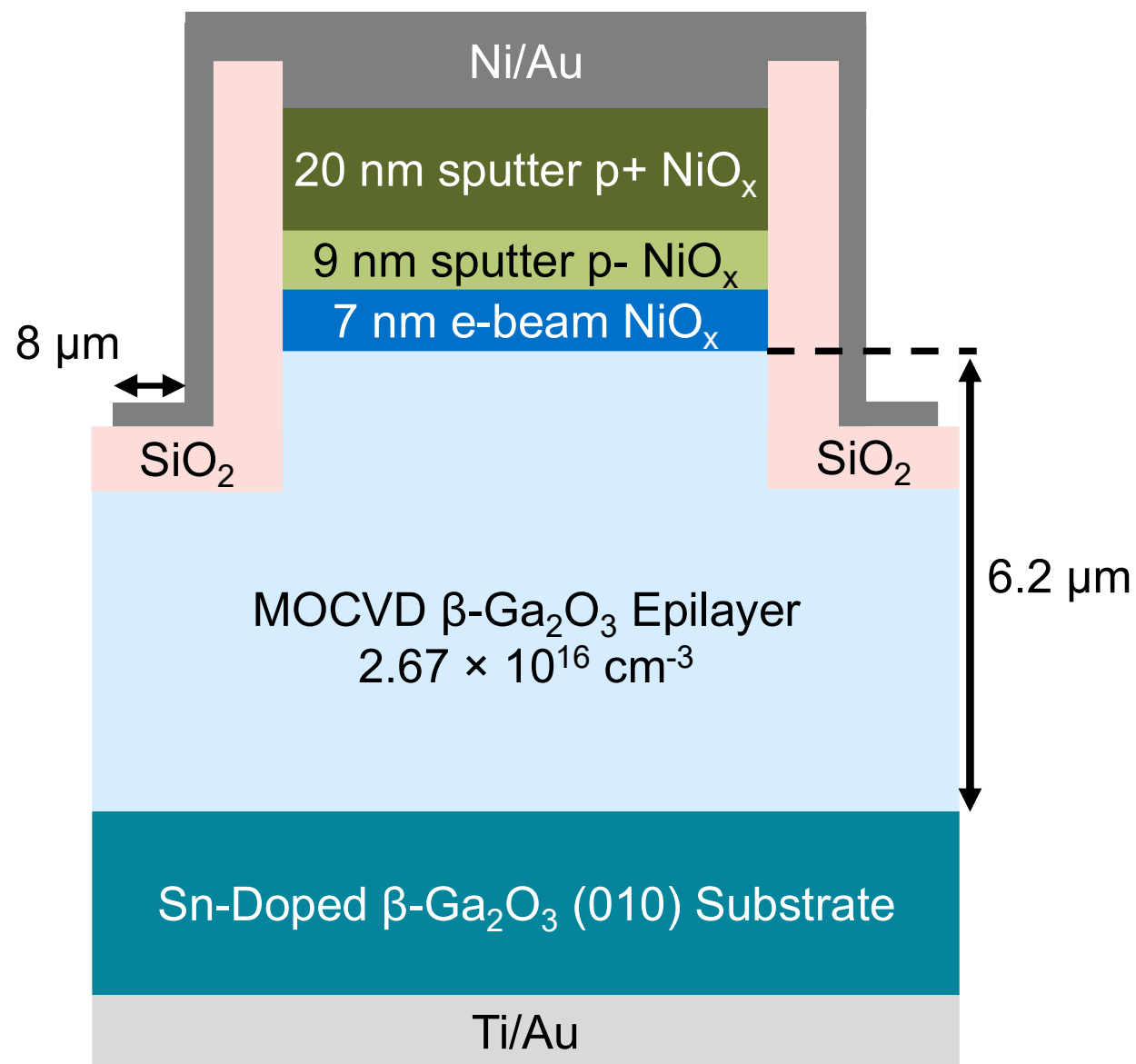

FIG. 1. Device cross-section schematic of the FP-HJD device fabricated on a 6.2 µm MOCVD-grown (010) $\beta$-$Ga_2O_3$ drift layer.

Analysis of the FP-HJDs was first done via high-voltage capacitance-voltage (C-V) measurements performed on a Keysight B1505A parameter analyzer. Measurements were performed on large area 1×1 $mm^2$ HJD devices to maximize the total capacitance and thus, measurement accuracy. The C-V curve in FIG. 2(a) shows the expected charge depletion with increasing voltage; however, the device broke down before reaching a constant capacitance value. Thus, punch-through could not be reached using the low damage FP-HJDs. The inset FIG. 2(b) shows the $1/C^2$ vs. V plot, with an extrapolated x-intercept giving a built in potential ($V_{bi}$) of 2 V, which is expected for $NiO_x$ HJDs.[57] Additionally, the net donor concentration ($N_D - N_A$) vs. depth profile was extracted from C-V and plotted as the blue curve in FIG. 2(c), showing a $N_D - N_A$ profile of ~2.67 × $10^{16}$ $cm^{-3}$ for the low-damage e-beam $NiO_x$ interlayer FP-HJDs. To reach punch-through and extract the epilayer thickness, sputtered HJD devices were fabricated without the protective e-beam interlayer to intentionally induce ion damage and deplete some of the charge. This depletion of charge allowed $N_D - N_A$ to be measured deeper into the epilayer without current leakage or device breakdown, as shown as the red curve in FIG. 2(c). Thus, we were able to determine an epilayer thickness of 6.2 µm. A critical observation in FIG. 2(c) is that there is no observable charge depletion in the low-damage FP-HJDs, therefore, the 7 nm e-beam interlayer successfully mitigated the ion damage from sputtering.[61] Additional analysis in the supplementary information FIG. S1 shows the low-damage e-beam interlayer HJDs having identical $N_D - N_A$ profiles to those of Schottky barrier diodes, confirming the efficacy of the e-beam $NiO_x$ interlayer in protecting the underlying (010) material from sputter ion damage. In addition to (010), we also found that the thin e-beam interlayer was successful in protecting the (011) sample surface from $NiO_x$ sputtering damage, as reported in Y. Liu *et al.*[60]

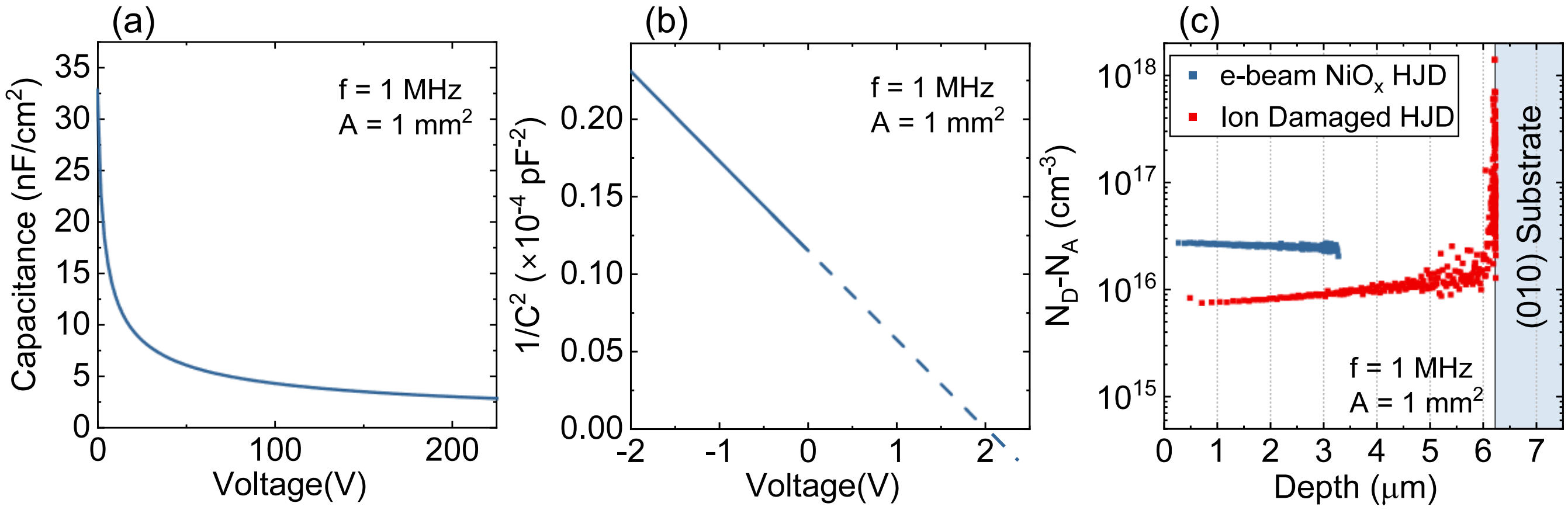

FIG. 2. (a) Capacitance vs. Voltage, (b) $1/C^2$ vs. V, and (c) $N_D - N_A$ vs. depth on a 1×1 $mm^2$ low damage e-beam $NiO_x$ FP-HJD on the 6.2 μm (010) MOCVD epilayer. The red dotted line in (c) is the $N_D - N_A$ vs. depth profile for an ion damaged 1×1 $mm^2$ $NiO_x$ FP-HJD to reach punch through.

Current density vs. voltage (J-V) measurements were performed using a Keithley 4200 parameter analyzer to determine the on-state and rectification performance of the FP-HJDs. Linear J-V measurements in FIG. 3(a) show a representative device with an on-current of 700 $A/cm^2$ at 4 V forward bias. FIG. 3(a) also depicts the differential specific on-resistance ($R_{on,sp}$) for the FP-HJDs, with a low minimum $R_{on,sp}$ of 2.36 $m\Omega\cdot cm^2$. In addition, a key observation in supplemental information FIG. S2 is that there is a minimal 1.4× increase in the $R_{on,sp}$ value for the e-beam interlayer FP-HJDs when compared to Schottky barrier diodes. The slight increase in $R_{on,sp}$ for the HJDs can likely be attributed to the added resistance of the $NiO_x$ layers. In contrast, for sputtered HJDs without an e-beam interlayer, the $R_{on,sp}$ was shown to increase by 9.4× when compared to Schottky diodes.[61] Thus, the minimal change in $R_{on,sp}$ observed in this work between the e-beam interlayer HJDs and SBDs further confirms the efficacy of the low-damage e-beam $NiO_x$ interlayer at protecting the (010) surface from sputter ion damage. Since the (010) material underneath the ICP isolation-etched region immediately surrounding the anode contact is expected to be depleted of charge[61] and thus, exhibiting a higher resistance than the epilayer underneath the anode metal, an assumption is made that there is minimal lateral current spreading in these devices. Therefore, only the anode area is used for normalization of devices. The rectification performance of the e-beam interlayer FP-HJDs was analyzed via a semi-log J-V plot (FIG. 3(b)) which shows a low reverse current leakage level at the noise floor of the tool at ~$5\times10^{-9}$ $A/cm^2$, leading to a rectification ratio of ~$10^{11}$. Additionally, the ideality factor was found to be 1.29, which is comparable to literature values for sputtered $NiO_x$ diodes.[53] An ideality factor between 1 and 2 potentially suggests some interface recombination current or other imperfections are present at the e-beam $NiO_x/Ga_2O_3$ interface.[65]

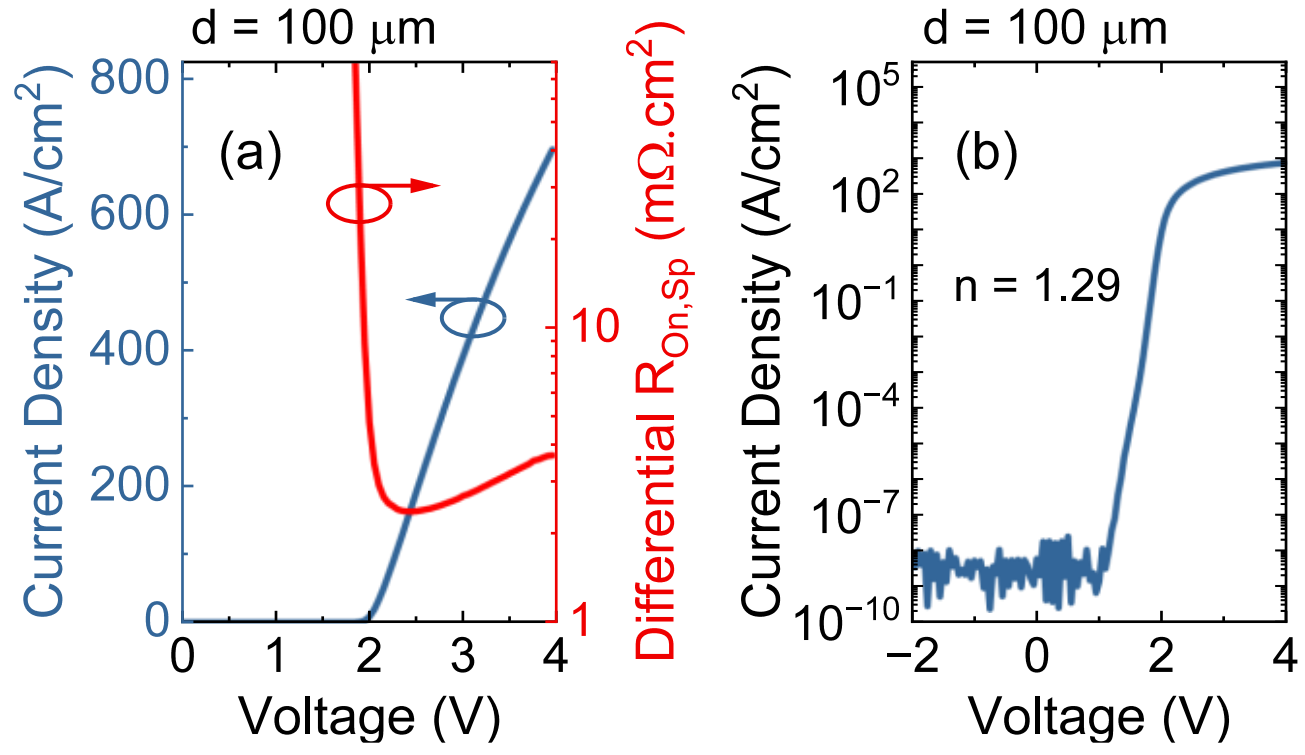

FIG. 3. (a) Differential $R_{on,sp}$, linear scale J-V, and (b) semi-log scale J-V characteristics of a representative 100 μm diameter e-beam interlayer $NiO_x$ FP-HJD on the 6.2 μm (010) MOCVD epilayer.

The reverse blocking capability of the FP-HJDs was analyzed via high-voltage J-V measurements performed on a Keysight B1505A. Before breakdown measurements were performed, the sample was submerged in Fluorinert FC-40 dielectric liquid to prevent electrical arcing. The breakdown voltage ($V_{BR}$) was then measured for both FP and non-FP HJDs, with the respective curves shown in FIG. 5. $V_{BR}$ on the non-FP devices was found to be 1.52 kV, whereas the FP devices $V_{BR}$ was 1.64 kV. Thus, a 120 V or 0.19 MV/cm improvement was observed for devices with the 1 μm $SiO_2$ FP dielectric field plate. This increase in breakdown voltage due to the field plate is not significantly large and signifies improvements can be made to the FP dielectric to achieve even greater device breakdown performance. The parallel plane electric field at breakdown ($E_{\|,max}$) under the anode center was calculated via a punch-through model and was found to be 3.83 and 4.02 MV/cm for the non-FP and FP-HJDs respectively, which is the highest $E_{\|,max}$ demonstrated for vertical diodes on (010) epilayers, as shown in FIG. 4(b).

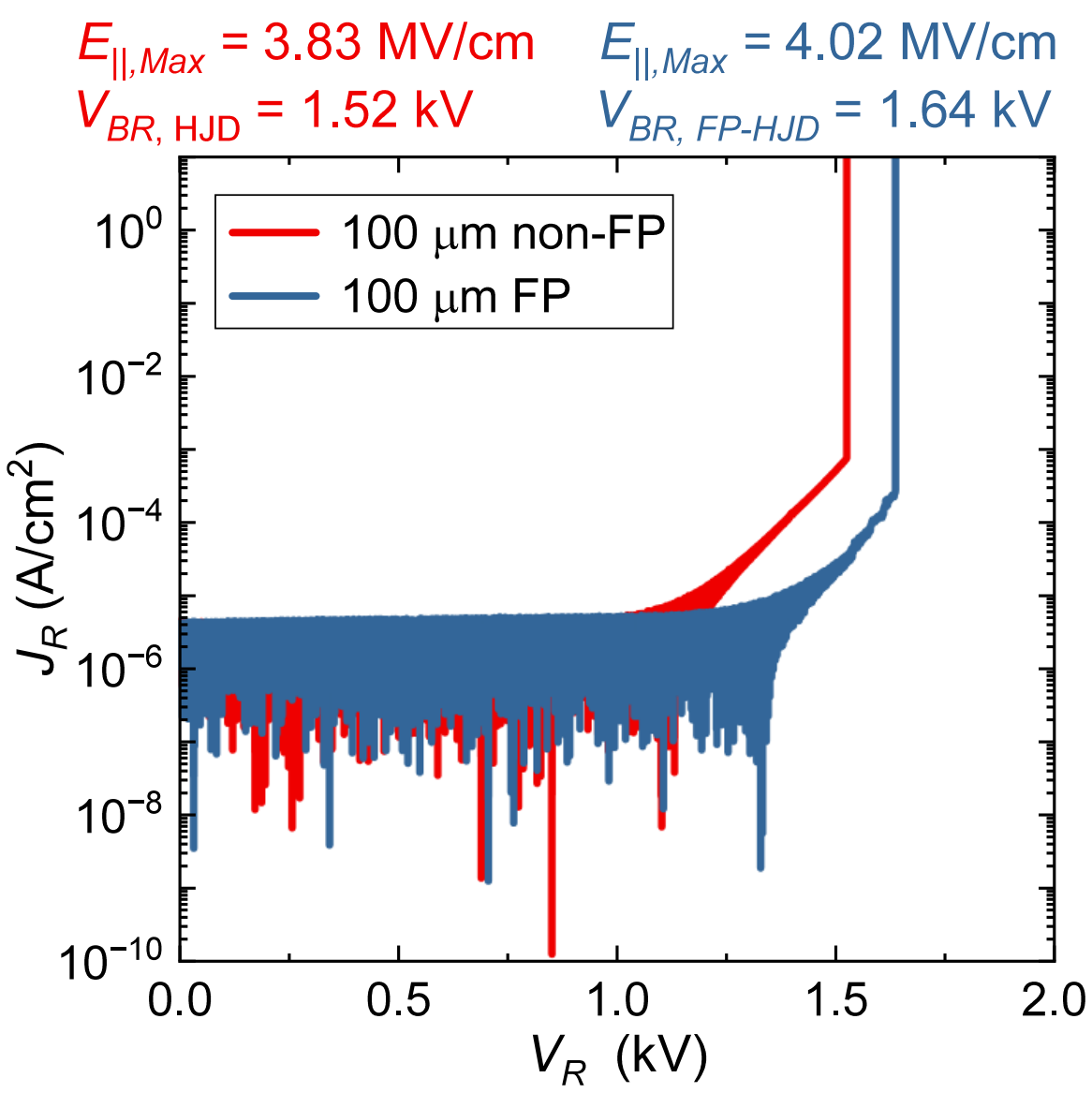

FIG. 5. Breakdown voltage vs. current density plot of both FP and non-FP HJD devices.

The power figure of merit (PFOM) was calculated by $V_{BR}^2/R_{on,sp}$ and was found to be 1.14 GW/cm$^2$. When compared to other MOCVD vertical diodes,[29,30,33,50,66–71] the PFOM of the $NiO_x$ HJDs in this work are among the highest reported, with record results among all (010) vertical devices, as depicted in FIG. 6(a). This confirms that the $NiO_x$ HJDs with a thin e-beam interlayer are highly effective in optimizing the $V_{BR}$ and $R_{on,sp}$ performance for diodes on (010) $\beta$-$Ga_2O_3$ material. Furthermore, an interesting observation in FIG. 7(b) is that on average, diodes fabricated on MOCVD-grown epilayers have lower parallel-plane breakdown electric field performance than the best results from HVPE-grown epilayers.[52–55,58,60,72–77] This highlights that there is still room for improvement on MOCVD device and epilayer design in the future. However, despite a significantly lower number of reported devices on MOCVD-grown epilayers, the comparable $E_{||,max}$ results between the two growth techniques show high promise for MOCVD-grown epilayers in the future.

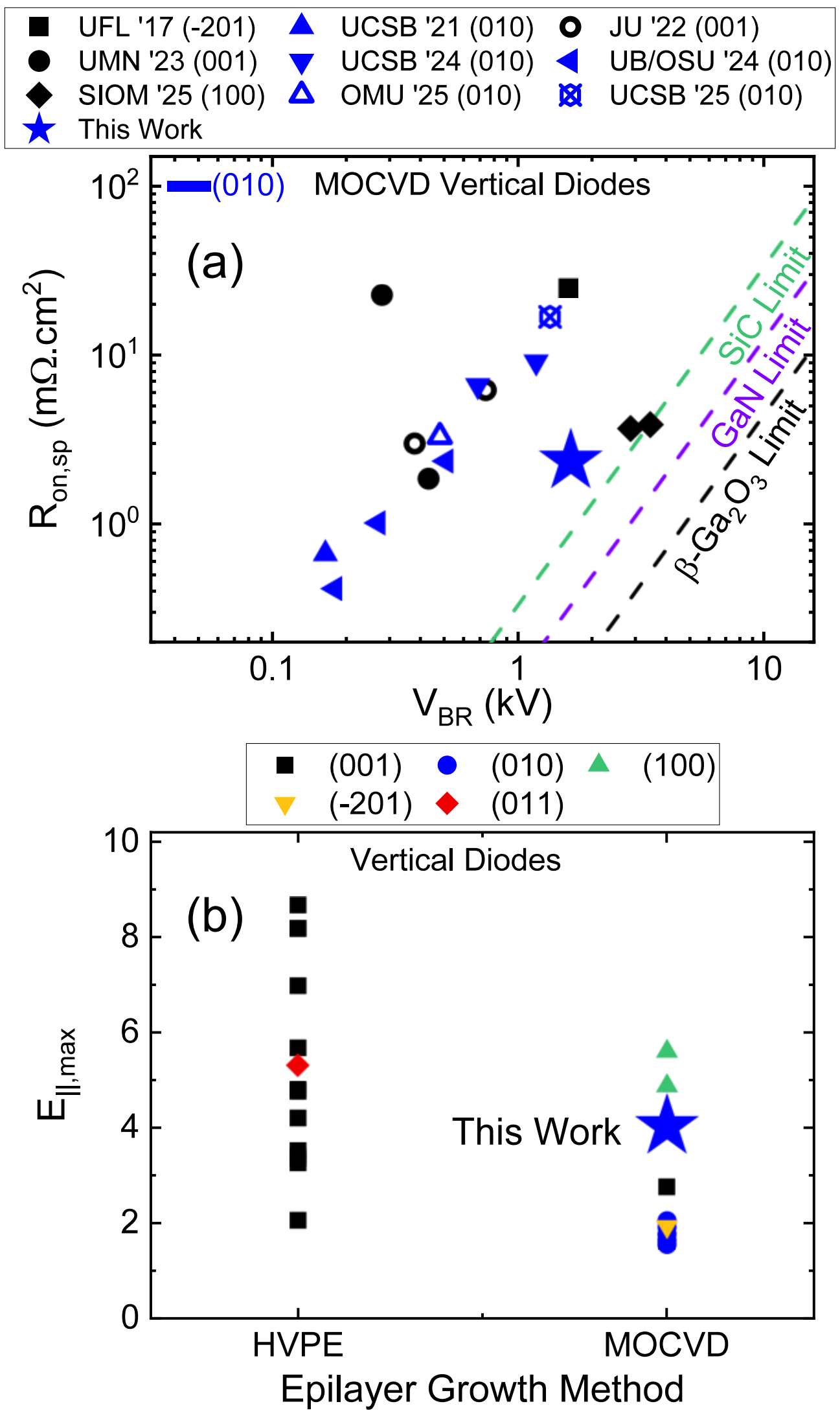

FIG. 8. Benchmarking plots for (a) breakdown voltage vs. on-resistance of vertical diodes on MOCVD-grown epilayers[29,30,33,50,66–71] and (b) comparison of parallel-plane electric field at breakdown for vertical diodes on both HVPE[52–55,58,60,72–77] and MOCVD-grown[27,29,30,33,50,66–71] epilayers.

In conclusion, we report on the use of a thin 7 nm low-damage e-beam deposited $NiO_x$ interlayer before $NiO_x$ sputtering to enable the creation of high-performance FP-HJD devices on (010) $\beta$-$Ga_2O_3$ films with low on-resistances, high $E_{||,max}$ values reaching 4.02 MV/cm, and a high PFOM of 1.14 GW/cm$^2$. FP-HJDs were fabricated on a 6.2 µm MOCVD epilayer with a $N_D - N_A$ of $2.67 \times 10^{16}$ cm$^{-3}$ grown on a conductive Sn-doped $\beta$-$Ga_2O_3$ (010) substrate. The important breakthrough for the devices was the utilization of a 7nm e-beam deposited $NiO_x$ interlayer before $NiO_x$ sputtering. Since sputtered $NiO_x$ causes severe damage in (010) epilayers,[61] the thin low-damage e-beam interlayer was successful in acting as a protective barrier and mitigating the sputter ion damage. The e-beam interlayer HJDs had high current densities comparable to SBDs of 700 A/cm$^2$ at 4 V, a $R_{on,sp}$ value of 2.36 mΩ·cm$^2$, HJD ideality factors of 1.29, $V_{bi}$ of 2 V, and rectification ratio of $10^{11}$. The FP-HJD devices on the (010) epilayers achieved breakdown voltages of 1.64 kV, leading to a $E_{||,max}$ of 4.02 MV/cm and a PFOM of 1.14 GW/cm$^2$, which are the highest values reported for vertical devices on (010) drift layers. These results mark a significant advancement for MOCVD-grown vertical $\beta$-$Ga_2O_3$ power devices and unlock a new pathway for fabricating high voltage HJDs on (010) $\beta$-$Ga_2O_3$ drift layers.

## SUPPLEMENTARY MATERIAL

The supplementary material contains additional data and discussions regarding the comparison of the net donor concentration and specific on-resistance values for the low-damage 7 nm $NiO_x$ e-beam interlayer field-plated heterojunction diodes with reference Ni/Au Schottky barrier diodes. These results show minimal change in net donor concentration and specific on resistance between the two devices.

## ACKNOWLEDGMENTS

The authors acknowledge funding from the ARPA-E ULTRAFAST program (DE-AR0001824) and Coherent / II-VI Foundation Block Gift Program. A portion of this work was performed at the UCSB Nanofabrication Facility, an open access laboratory.

## AUTHOR DECLARATIONS

The authors have no conflict to disclose.

## DATA AVAILABILITY

The data that support the findings of this study are available from the corresponding author upon reasonable request.

## Supplementary Material

# 4 MV/cm (010) $\beta$-$Ga_2O_3$ Heterojunction Diodes Realized by Low-Damage e-Beam $NiO_x$ Interlayers

Carl Peterson,[1,a)] Yizheng Liu,[1] Chinmoy Nath Saha,[1] Rachel Kahler,[1] Akhila Mattapalli,[1] and Sriram Krishnamoorthy[1,a)]

[1] *Materials Department, University of California Santa Barbara, Santa Barbara, California, 93106, USA*

[b)] Author to whom correspondence should be addressed. Electronic mail: carlpeterson@ucsb.edu and sriramkrishnamoorthy@ucsb.edu

### S1. Comparison of $N_D$-$N_A$ for e-beam interlayer heterojunction diodes and reference Schottky diodes

FIG. S1(a) shows the cross-section schematic of both the circular e-beam interlayer heterojunction diode (HJD) and reference Schottky barrier diode (SBD). These devices were co-fabricated on an 8 $\mu$m (010) $\beta$-$Ga_2O_3$ MOCVD-grown epitaxial drift layer. The homoepitaxial growth was performed on 5x5 $mm^2$ Sn-doped conductive (010) $\beta$-$Ga_2O_3$ substrate with no intentional miscut commercially acquired from NCT. Growth was performed with identical precursors (TMGa, $O_2$, $SiH_4$, Ar carrier gas) and growth conditions (1000 ºC, 15 Torr. VI/III ratio of 523, 3.7 $\mu$m/hr.) as the epilayer in the main manuscript. Fabrication of the HJD devices was also identical to that in the main manuscript, except there was no junction isolation or field plating. Processing starting with an annealed Ti/Au Ohmic backside contact. The anode stack consisted of a thin 7 nm $NiO_x$ layer deposited via e-beam to protect the surface from sputter damage, a self-aligned reactively sputtered 9 nm $p^-$ $NiO_x$/20 nm $p^+$ $NiO_x$ stack using a Ni target and a 150 W $O_2$/Ar plasma, and finally a self-aligned Ni/Au metal stack deposited via e-beam evaporation. SBDs were simply created via e-beam Ni/Au deposition with lithographic patterning and a lift-off method.

Capacitance-voltage (C-V) measurements were taken via a Keithley 4200 parameter analyzer and the extracted $N_D$-$N_A$ vs. depth results are shown in FIG. S1(b). The key observation from these results is that there is no observable difference between the $N_D$-$N_A$ concentrations of the HJD and SBD devices. In contrast, literature results show an 85% reduction in zero-bias $N_D$-$N_A$ for sputter-only $NiO_x$ HJDs on (010) $\beta$-$Ga_2O_3$ epilayers without an e-beam interlayer.[1] This solidifies that the 7 nm e-beam $NiO_x$ interlayer is highly effective at preventing charge depletion due to ion damage from the subsequent $NiO_x$ sputtering.

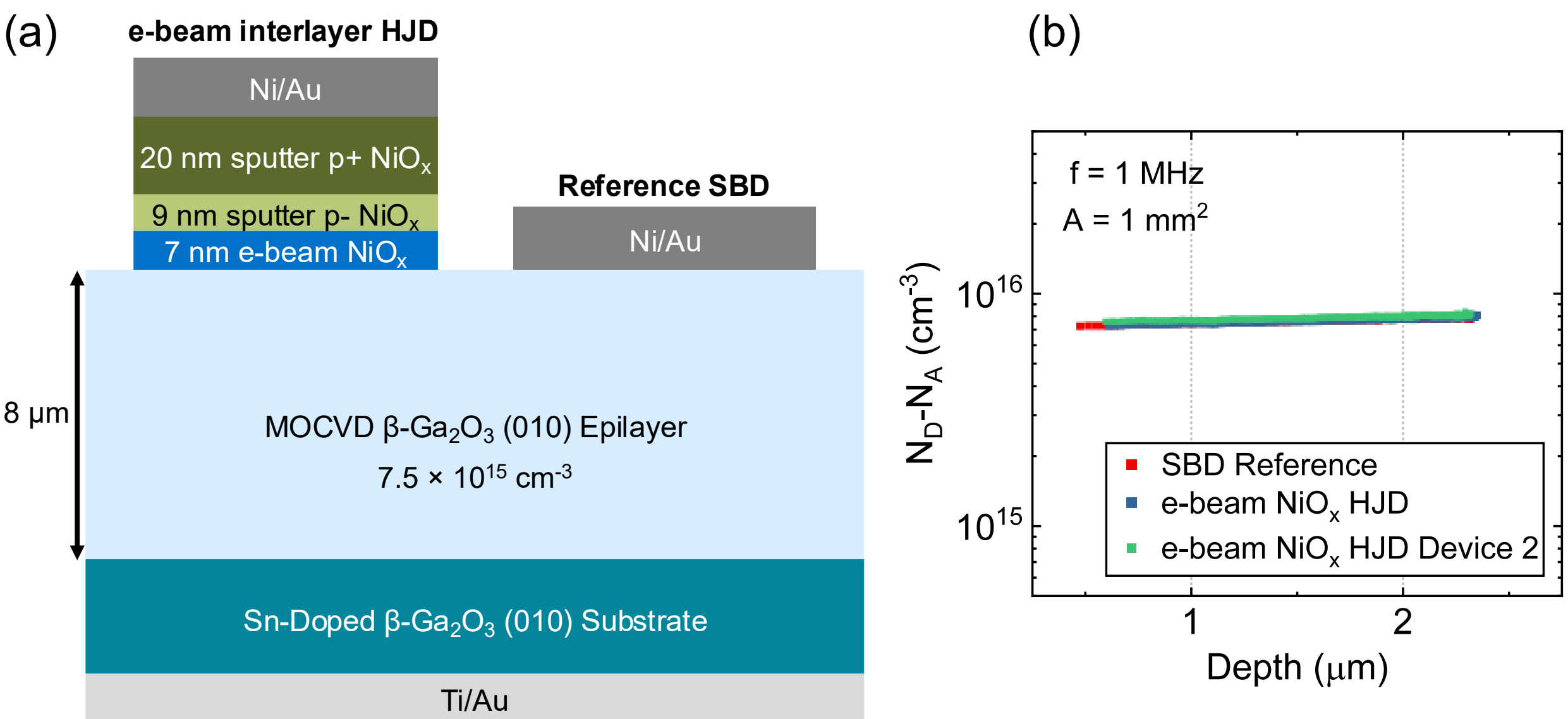

*FIG. S1. (a) Cross-section schematics of the e-beam interlayer HJDs with co-fabricated reference Ni/Au SBDs on a (010) β-$Ga_2O_3$ epilayer. (b) $N_D$-$N_A$ vs. depth profiles for the e-beam interlayer HJDs and a reference SBD extracted from C-V measurements, showing identical profiles between the two structures.*

## S2. Comparison of specific on-resistance ($R_{on,sp}$) for e-beam interlayer heterojunction diodes and reference Schottky diodes

For the e-beam interlayer HJD and reference SBD devices shown in FIG. S1(a), current density vs. voltage (J-V) measurements were taken on a Keithley 4200 parameter analyzer. FIG. S2(a) shows the linear J-V characteristics, with both the HJD and SBD devices reaching current densities of >400 A/cm$^2$ which is good for power rated devices. The log-scale J-V in FIG. S2(b) shows a high rectification ratio of ~$10^{10}$ and tool noise-floor limited leakage current levels for both devices. The ideality factor for the HJD was shown to be slightly higher than the reference SBD, with a value of 1.29 and 1.01 respectively. Most importantly, the comparison of the $R_{on,sp}$ values are shown in FIG. S2(c), with the e-beam interlayer HJD and SBD having minimum values of 3.87 mΩ.cm$^2$ and 2.70 mΩ.cm$^2$ respectively. This corresponds to a 1.43× increase in the $R_{on,sp}$ between the SBD and HJD, which is quite minimal compared to the 9.4× increase in $R_{on,sp}$ observed in sputtered $NiO_x$ HJDs without the e-beam $NiO_x$ interalayer.[1] The minimal change in $R_{on,sp}$ observed further confirms that the e-beam $NiO_x$ interlayer was successful in mitigating ion damage from the sputter $NiO_x$ process.

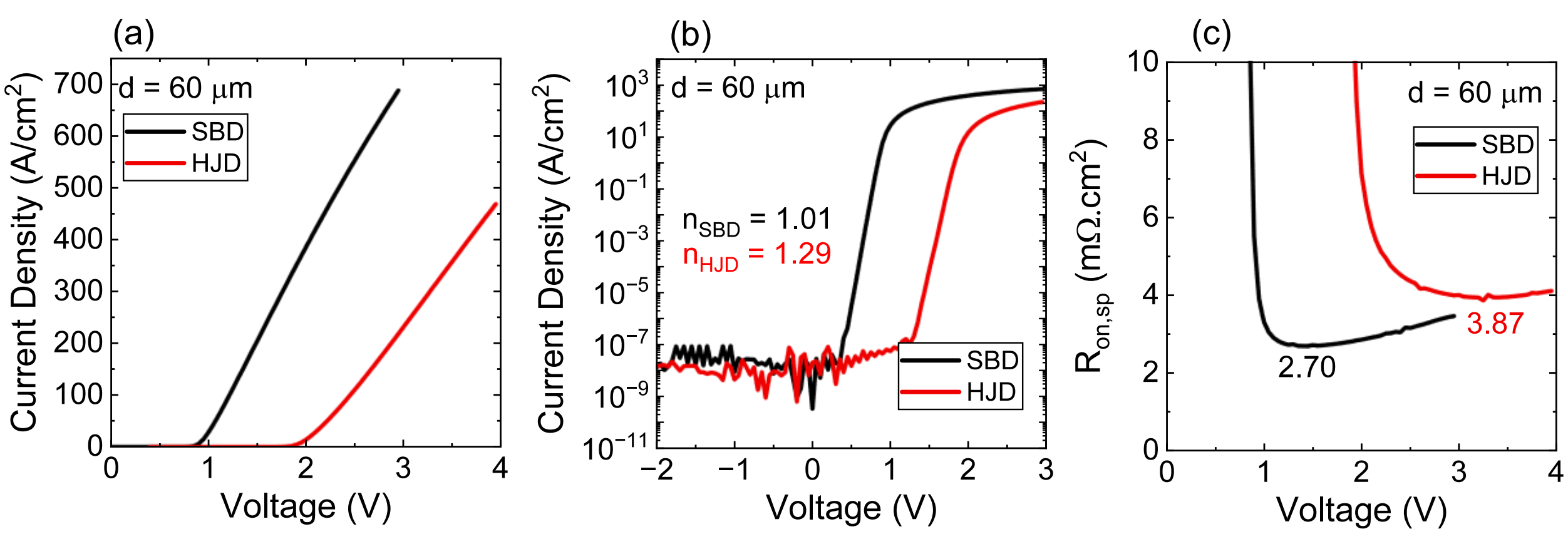

*FIG. S2. (a) Linear J-V, (b) Semi-log-scale J-V, and (c) extracted differential $R_{on,sp}$ curves for both the e-beam interlayer HJDs and co-fabricated reference Ni/Au SBDs on a (010) β-$Ga_2O_3$ epilayer. Results show a minimal 1.43× change in the $R_{on,sp}$.*